# Landau equations and asymptotic operation

Fyodor V. Tkachov

Institute for Nuclear Research of the Russian Academy of Sciences,
Moscow 117312, RUSSIA

*Abstract* The pinched/nonpinched classification of intersections of causal singularities of propagators in Minkowski space is reconsidered in the context of the theory of asymptotic operation as a first step towards extension of the latter to non-Euclidean asymptotic regimes. A highly visual distribution-theoretic technique of singular wave fronts is taylored to the needs of the theory of Feynman diagrams. Besides a simple derivation of the usual Landau equations in the case of the conventional singularities, the technique naturally extends to other types of singularities e.g. due to linear denominators in non-covariant gauges etc. As another application, the results of Euclidean asymptotic operation are extended to a class of quasi-Euclidean asymptotic regimes in Minkowski space.

## Introduction    1

The problem of constructing asymptotic expansions of multiloop Feynman diagrams in masses and momenta for various asymptotic regimes in a phenomenologically meaningful and calculationally useful form is the key analytical problem of applied Quantum Field Theory. This is because some sort of asymptotic expansion seems to always be involved in an important way — explicitly or implicitly — whenever one deals with Feynman diagrams in applications (see [1] for a systematic discussion).

The method of *asymptotic operation* (*As-operation* for short; for a review see [1]) has proved to be the most powerful one for solving the asymptotic expansion problem within applied QFT. It originated [2]–[5] in the context of studies of analytical calculational methods for the coefficient functions of short-distance operator product expansion. The resulting algorithms [3], [6] (extended to arbitrary Euclidean expansions in [5]) allow one to obtain contributions to coefficient functions from individual diagrams in a maximally simple form, which comes about as a natural consequence of a property of *perfect factorization* of the expansions obtained with *As*-operation [3], [4]. Due to the perfect factorization, calculations of coefficient functions are much simplified and are directly reduced to integrals for which there exist powerful algebraic algorithms [7]. As a result, a class of next-next-to-leading order calculations had been reduced to feeding the corresponding diagrams into a computer (e.g. [8]–[11]). A regularization-independent analysis [12], [13] demonstrated a remarkable compactness of formal proofs within the method of *As*-operation and established the latter as a superior alternative to the traditional (BPHZ-type) methods in the theory of Feynman diagrams (for a systematic dicussion of the differences between the two paradigms see [1]).

It became clear quite early that the method of asymptotic operation easily extends to the most general Euclidean asymptotic regimes [4] and that simplicity of the resulting formulas remains its characteristic feature [5]. Moreover, the general philosophy of the method of *As*-operation [4], [1] is by no means specific to Euclidean problems (first examples of non-trivial application of the method in essentially non-Euclidean situations can be found in [14], [15]).

The purpose of the present paper is to begin a systematic extension of the theory of asymptotic operation to non-Euclidean asymptotic regimes.

### Asymptotic operation and causal singularities    1.1

A necessary preliminary step is to reinterpret the well-known pinch/non-pinch classification of intersections of light-cone and mass-shell singularities of causal propagators in purely distribution-theoretic terms as is necessary within the framework of *As*-operation. Indeed, the first question one is confronted with is as follows.

The formulas of *As*-operation for Euclidean asymptotic regimes were derived (cf. [4]) using the Wick rotation so that one dealt with Euclidean momenta throughout the analytical part of the derivation. However, the final formulas are valid irrespective of the Wick rotation used at intermediate steps. But then it should be possible to derive the Euclidean *As*-operation directly in Minkowski space.

The technical problem one encounters here is due to the more involved structure of singularities of Feynman propagators in Minkowski space as compared with the Euclidean situation. Indeed, the Euclidean scalar propagator has the form $(p^2 + m^2)^{-1}$ where $p^2 > 0$ for all $p \neq 0$. When formally expanded in powers of $m$, this results in singularities at the *isolated point* $p^\mu = 0$. If there are other integration momenta, then the singularity of a single propagator is localized on a linear manifold. On the other hand, in a Minkowskian situation the standard scalar propagator becomes $(-p^2 + m^2 - i0)^{-1}$ where $p^2$ can be both positive and negative so that there is a singularity localized on the *non-linear manifold* described by the second-order algebraic equation $p^2 = m^2$. If $m$ is one of the expansion parameters ($m = O(\kappa)$), then the expanded expression has singularities on the light cone $p^2 = 0$. Moreover, the point $p = 0$ is a singular point of the light cone manifold in the sense of differential geometry and should be treated separately. The singularity at $p = 0$ is traditionally referred to as soft singularity[i]. Other singular points have the property that the propagator near each such point becomes a well-known Sohotsky distribution, namely, $(x - i0)^{-1}$ where $x$ is a properly chosen one-dimensional local coordinate. To emphasize their analytical nature, such singularities can be referred to as *Sohotsky-type singularities*. To emphasize their physical origin, the term *causal singularities* can be empoyed. While the causal $-i0$ prescription ensures that the propagator itself is a well-defined distribution everywhere in the space of $p$, this is not necessarily true for products of such distributions. The causal Sohotsky-type singularities may overlap and make the product non-integrable by power counting.

In general, singularities that are non-integrable by power counting result in non-trivial contributions to asymptotic expansions. Therefore, to obtain a Minkowski-space derivation of the Euclidean *As*-operation one has to understand why the intersections of causal singularities do not contribute in the case of Euclidean asymptotic regimes.

The usual explanation of non-contribution of causal singularities in Euclidean case is, because the energy integration contours are not pinched and one can deform them away from such singularities (cf. the derivation of the Landau equations that give a criterion of occurence of pinches [16], [17]). But this is the same as to say that the Wick rotation can be performed, and cannot satisfy us. Indeed, eventually one will have to answer the question of what happens near the points that separate pinched and non-pinched singularities.[ii] Then one would have to cope with the fact that the latter are disposed of via contour deformations while no such trick is allowed for the former. To avoid such a situation one has to reinterpret the mechanism of non-contribution of non-pinched singularities in a manner that avoids contour deformations.

Moreover, the technique of *As*-operation is, strictly speaking, not naturally compatible with contour deformations. Indeed, whenever distributions are involved one has to deal with test functions which in the theory of *As*-operation should be localizable — i.e. zero everywhere except a small neighbourhood of a point which one studies. This, however, cannot be achieved with analytic functions, so that analytic continuation is no longer allowed.

Lastly, as was stressed e.g. in [17] (Chapter 2 and refs. therein), an accurate derivation of the Landau equations via contour deformations in a general case requires a use of the notoriously cumbersome methods of algebraic topology. This is unsatisfactory because a calculationist would always prefer a direct analytical argument at a "microlocal" level — and argument that could be directly connected with how one calculates, say, asymptotic expansions — to an indirect one such as the one based on contour deformations.

Wave fronts and Feynman diagrams 1.2

An alternative technique that would allow a direct study of overlapping causal singularities is based on the notion of *singular wave front* of a distribution (or simply *wave front*) introduced by Hörmander [18] and Sato [19] in the context of studies of propagation of singularities of generalized solutions of partial differential equations[iii]. Whereas Hörmander worked with distributions proper [20], Sato considered a related notion of hyperfunctions [21]. From a practical point of view, however, the two definitions are equivalent (cf. the review [22]). The wave front describes singularities of a given function (their positions and "orientation", cf. Sec. 4.23 below; for a general definition see [22], [23]). The key technical result of interest to us here is a criterion of existence of products of distributions in terms of their wave fronts (described e.g. in [23], Theorem IX.45 essentially due Hörmander). However, we will not need the criterion in full generality, and in Sec. 4 its simplified version will be taylored to the needs of the theory of Feynman diagrams.

Sato et al. [24] were first to apply the wave-front-related techniques to Feynman diagrams. They did so from the point of view of the theory of hyperfunctions [21], which was rather natural because hyperfunctions can be roughly thought of as boundary values of analytic functions.[iv] The main result of [24] is a description of non-analyticities of Feynman diagrams in terms of certain closed formulas (to be compared with the implicit description via the Landau equations). Despite being in a formal sense explicit, such formulas nevertheless seem to be less useful in physical applications than, say, the transparent interpretation of pinched singularities due to Coleman and Norton [26].[v, vi, vii] Yet the notion of wave front — if used properly, i.e. without excessive generality — is a handy tool in the theory of Feynman diagrams due to its visual simplicity.

---

[i] By association with soft (zero-momentum) photons in QED; analytically, this is the hardest singularity generated by a propagator.

[ii] I first heard this question explicitly stated from G. Sterman.

[iii] Cf. the light cone singularities of causal propagators that are special solutions of the wave equation in coordinate representation.

[iv] The relevant physically significant results include e.g. the edge of the wedge theorem in the context of studies of dispersion relations (see e.g. [25]).

[v] I than J. Collins for a discussion of the Coleman and Norton interpretation.

[vi] The work of Sato et al. should not be judged too narrowly: they studied a new "mechanism of proofs" in an applications-motivated problem in exactly the same way as physicists would design an experiment to test, say, a particle production mechanism — irrespective of its immediate usefulness.

[vii] The results of [24] were later reproduced in a series of publications by Smirnov (see his compilation [27] and refs. therein) who claimed to have used a "different" technique of "singular wave front" of Hörmander rather than the "analytical wave front" of Sato et al. Such claims cannot be sustained in view of the exact equivalence of the two versions of the definition of wave front in the context of Feynman diagrams, as was emphasized in connection with Smirnov's "results" in [28].

### Purpose 1.3

It was already emphasized elsewhere [1] that the mathematical apparatus of asymptotic operation includes a collection of techniques that cannot be found in any single source on distribution theory.[i] This is why it would be useful to present an account of the corresponding techniques in a form taylored to the specific needs of theorists that have to deal with Feynman diagrams in applied problems. The goal of the present paper from a technical point of view is to do that with respect to the technique of wave fronts.

In this respect note, first, that applications to Feynman diagrams do not require the use of wave fronts in full generality. Second, whereas the usual treatments consider "static" products of singular functions, we wish to study asymptotic expansions so that not only existence of products has to be established but also the fact that they possess the required asymptotic approximation properties. All the necessary mathematical apparatus will be explained in detail.

From the point of view of the theory of multiloop diagrams, we are going to: rederive the Landau equations [16], [17] in a purely distribution-theoretic fashion, i.e. avoiding the use of contour deformations as is appropriate within the theory of *As*-operation; demonstrate how the technique can be applied to analysis of singular functions of a more general form than the standard causal propagators (cf. the problem of singularities in non-covariant gauges [31]); extend the Euclidean theory of asymptotic operation [4], [5] to the simplest class of non-Euclidean asymptotic regimes (the so-called quasi-Euclidean ones).

It should be emphasized that the motivations of this work go somewhat beyond providing a "better derivation" of the Landau equations. The problem of asymptotic expansions of multiloop diagrams in non-Euclidean regimes is a notoriously complex one, and complex problems are solved by building an adequate intuition based on — and together with — and adequate formalism. So our aim is to present a more flexible technique than the one based on contour deformations and to reconsider the classification of singularities in a manner suitable for a subsequent construction of non-Euclidean infrared *R*- and *As*-operations. The fact that the Landau equations reemerge as a result is in principle of little consequence.

### Plan 1.4

The plan of the paper is as follows. In the descriptive Sec. 2 we establish the context in which to study non-Euclidean *As*-operation. First the systematic notations are summarized to describe the products of singular functions we wish to deal with. Then a summary of *As*-operation is presented emphasizing the point that its structure is independent of whether one works in Euclidean or Minkowski space. Further, we identify the nonlinear causal singularities and briefly discuss the distinctively non-Euclidean geometrical phenomenon of osculating singularities. The conventional treatments avoid discussing osculating singularities at a microlocal level altogether, but the issue cannot be circumvented if one aims (as we do in the theory of *As*-operation) at a complete solution of the asymptotic expansion problem.

---
[i] The original source on the theory of distributions is [20]. Good textbooks are [29] (with many examples to different physical problems) and [30] (with proofs of the general results of the theory of distributions that avoid the abstract theory of topological spaces). Wave fronts are briefly discussed in [23].

Because the method of *As*-operation iteratively reduces studying any however complex geometrical patterns of singularities to the case of singularities localized at isolated points, Sec. 3 contains an exhaustive treatment of one-dimensional Sohotsky distributions, the one-dimensional prototypes of causal singularities. We consider in detail the issue of when products of such singularities exist despite divergence by power counting. Then this fact is reinterpreted in terms of Fourier transform thus preparing ground for an extension to many dimensions. Finally, a simple extension to asymptotic expansions of products of Sohotsky distributions is presented.

Sec. 4 treats the multidimensional case. First, general intersections of Sohotsky-type singularities are considered, and a special version of the highly visual notion of singular wave front is deduced in a few easy steps, along with the necessary heuristic motivations. A criterion of local existence of such products (a special case of the more abstract one due to Hörmander) is derived as an obvious generalization of the one-dimensional results, both in an analytical and geometrical form. Then it is shown that the same criterion holds for the case of osculating singularities; at this point the heuristic power of the notion of wave front becomes apparent.

Finally, in Sec. 5 we turn to multiloop diagrams and present a one-step derivation of the Landau equations. Then we briefly discuss how the developed formalism helps to understand the structure of singularities in the theory of non-covariant gauges, and how the results of Euclidean theory of asymptotic operation are extended to the class of quasi-Euclidean asymptotic regimes.

## Causal singularities and asymptotic operation 2

### Notations for integrands 2.1

Our notations will generally agree with [4]. Let us briefly summarize them together with whatever extensions are needed in the non-Euclidean case. This will also introduce one into the context of the problem.

The object one deals with is $G(p)$, the integrand of a multiloop diagram $G$, whereas $p = (p_1, \ldots p_l)$ is the collection of all its integration (loop) momenta: $dp = d^D p_1 d^D p_2 \ldots$. The space of $p$ is denoted as $P$. The concrete value of $D$ (the number of scalar components of each $p_i$) is not important. It is often convenient to ignore the structure of $P$ altogether and to consider it as an abstract vector space of finite dimensionality. The diagrammatic interpretation of $G$ will not be important either and we will treat $G(p)$ simply as a product of singular factors. Its structure is described below.

A general scheme is that one starts with $G(p)$ and constructs a distribution from it. The conditions governing such a construction are always local. This means one should first perform such a construction for all sufficiently small $\mathcal{O}$ (neighbourhoods of singular points). After that the transition to a distribution defined on the entire $P$ is achieved via the standard trick of the decomposition of unit [29], [30]. For this reason, we focus attention only on what happens in a small open region $\mathcal{O}$.

Furthermore, we will ignore the polynomial factors in the numerator of $G(p)$ that are due to interaction vertices or non-scalar particles. This is because the *As*-operation, by definition, constructs a complete asymptotic expansion in powers and logarithms of the expansion parameter. Such expansions are unique and, therefore, commute with multiplications by polynomials [4]. This is a general property that remains valid in the non-Euclidean case. Therefore, the presence of polynomial factors is of no significance as far as the analytical mechanism of *As*-operation is concerned. They play no role in the study of the geometry of causal singularities and will be ignored.

So, the products of singular factors that we consider in this paper have the following form

$$G(p) = \prod_{g \in G} g(p), \qquad 2.2$$

where

$$g(p) \equiv \Delta_g(p) \equiv \left(D_g(p) - i0\right)^{-n_g}. \qquad 2.3$$

$D_g(p)$ is whatever expression happens to occur in the denominator of the $g$-th factor. Eq. 2.3 fixes the relative sign of $D_g(p)$ with respect to the infinitesimal imaginary part. In this work we consider only such regions $\mathcal{O} \subset P$ that the change of sign does not occur within $\mathcal{O}$ in all factors in the product.

### Structure of denominators 2.4

In the case of the standard scalar propagator,

$$D_g(p) = -l_g^2(p) + m_g^2, \qquad 2.5$$

where $m_g$ may or may not be zero. The quantity $l_g(p)$ is the momentum flowing through the $g$-th line of $G$; it is a linear combination of $p_i$ and the external momenta. The concrete form of $l_g$ is determined by the topology of the diagram and the pattern of its external momenta. When studying the general analytical mechanisms, we make assumptions neither about the topology, nor about the pattern of external momenta (i.e. whether they are put on mass shell, taken at zero values etc.).

The construction of *As*-operation requires to perform various expansions of the factors [1], [4]. This explains why we allow the powers $n_g$ to differ from 1. Moreover, in the non-Euclidean case, such expansions may change the form of denominators. In particular, there can emerge denominators with a linear dependence on the momentum,

$$D_g(p) = n_g p + m_g^2, \qquad 2.6$$

where $n_g$ is a parameter vector, etc. The examples where such factors occur are: the string operators in the context of the Sudakov problem etc. [32]; the propagator of the heavy quark in the effective heavy quark theory (cf. e.g. the effective QCD Lagrangian in [33]). The non-covariant gauges [34] also involve factors with denominators similar to 2.6 but with the sign of the imaginary part different in different regions of $P$.

In view of the variety of possible denominators, it makes sense to keep irrelevant details out of the way and simply regard $D_g(p)$ as a smooth function of $p$.

The expansion parameter is denoted as $\kappa$. The dependence on $\kappa$ of various quantities is indicated, as usual [4], with an additional argument, e.g. $G(p,\kappa)$, $D_g(p,\kappa)$ etc. Its particular form is determined by the specific problem and the asymptotic regime chosen.

### General structure of the asymptotic operation 2.7

It is important to understand that the general structure of the *As*-operation is the same irrespective of whether one deals with Euclidean or non-Euclidean problems. This is because neither the motivations for the distribution-theoretic point of view presented in Sec. 4 of [4], nor the extension principle of Sec. 6 of [4], nor even the general structure of the arguments of Sec. 11 of [4] which established the structure of *As*-operation, depend on that assumption. Let us summarize the prescriptions of *As*-operation emphasizing this fact.

The asymptotic operation $\mathsf{As}_\kappa$ (with respect to a concrete small parameter $\kappa$ which implies a concrete choice of the asymptotic regime) is meant to be applied to a product of singular factors $G(p,\kappa)$ regarded as a distribution in the integration momenta $p$. It is then supposed to yield an asymptotic expansion of $G(p,\kappa)$ in powers and logarithms[i] of $\kappa$ in the sense of distributions. At an operational level, the construction involves the following steps:

(i) Applying the formal Taylor expansion $\mathsf{T}_\kappa$ in powers of $\kappa$ to each factor $\Delta_g(p,\kappa)$ in the product,

$$G(p,\kappa) \equiv \prod_{g \in G} g(p,\kappa)$$
$$\to \mathsf{T}_\kappa \circ \prod_{g \in G} g(p,\kappa) \equiv \prod_{g \in G} \mathsf{T}_\kappa \circ g(p,\kappa), \qquad 2.8$$

which results in a formal series in powers of $\kappa$, identification of potentially non-integrable singularities in the resulting expression and their geometrical classification. Here one determines the number and geometric structure of the manifolds in $P$ on which those singularities are localized. One here deals with an hierarchy of intersecting manifolds of various dimensionalities. Note that all such manifolds occur as intersections of singular manifolds for individual factors $g(p,\kappa)$.

(ii) Determining which factors from the product contribute at each singular point. Here one establishes a correspondence between the singular manifolds $\pi_\gamma$, the subproducts $\gamma \subset G$ consisting of factors that are singular there, and the diagrammatic interpretation of such subproducts as some kind of subgraphs of the initial Feynman graph.[ii]

(iii) Studying the analytical structure of the singularities (some form of power counting etc.) and constructing a subtraction procedure that transforms the singular and, in general, non-integrable formal expansion into a correctly defined distribution on the entire $P$.[iii] As in [13], we denote it as $\tilde{\mathsf{R}}$:

---

[i] The powers need not be integer, e.g. square roots are allowed etc. We are talking about "powers and logarithms" although the analytical fact that the expansions we deal with run in powers and logarithms of the expansion parameter and do not contain, say, $\log \log \kappa$ terms, cannot be decided a priori. However, such an assumption is not used in the actual construction of *As*-operation and its correctness is established along the way. Note also that proving the power-and-log structure of expansions is a much simpler problem than establishing their precise diagrammatic form. References to some such studies can be found e.g. in [4].

[ii] Note that the resulting diagrammatic images are determined by the underlying analytical structures (in particular, the asymptotic regime) and need not fit exactly into the standard categories of the theory of graphs. In fact, a graph-theoretic characterization, even if possible, may be too cumbersome to be useful, or simply irrelevant (except in a very special context, e.g. when rewriting known results in a "rigorous" form; cf. [27]).

[iii] In certain cases one may do without such a subtraction procedure if a convenient regularization is available (the dimensional regularization is

$$\prod_{g \in G} \mathsf{T} \circ g(p,\kappa) \to \widetilde{\mathsf{R}} \circ \prod_{g \in G} \mathsf{T} \circ g(p,\kappa) . \qquad 2.9$$

The subtraction procedure $\widetilde{\mathsf{R}}$ should be minimal in that it should not modify the leading power behaviour near singular points. It is analytically somewhat similar to the Bogoliubov $R$-operation in coordinate representation.

(iv) Constructing the counterterms $\mathsf{E}_\gamma(p,\kappa)$ localized at the singular manifolds $\pi_\gamma$ (thus fixing the finite arbitrariness in the definition of $\widetilde{\mathsf{R}}$ ):

$$\widetilde{\mathsf{R}} \circ \prod_{g \in G} \mathsf{T}_\kappa \circ g(p,\kappa)$$
$$\to \widetilde{\mathsf{R}} \circ \prod_{g \in G \setminus \gamma} \mathsf{T}_\kappa \circ g(p,\kappa) \times \mathsf{E}_\gamma(p,\kappa) \equiv \mathsf{As}_\kappa \circ \prod_{g \in G} g(p,\kappa) . \qquad 2.10$$

The inclusion of $\mathsf{E}_\gamma(p,\kappa)$ is necessary in order to transform the expression into a correct expansion in the sense of distributions. $\mathsf{E}_\gamma(p,\kappa)$ contain all the non-analytic (logarithmic, square root etc.) dependence on $\kappa$. They are localized on the singular manifolds $\pi_\gamma$ and, therefore, are linear combinations of $\delta$-functions. The number of derivatives of the $\delta$-functions is determined by the power counting at the previous step, whereas the non-trivial coefficients are found using the consistency conditions [4]. The coefficients of the counterterms of the $As$-operation are detemined uniquely [4].

Note that one needs some form of explicit parametrization of $\pi_\gamma$ to write down the structure of $\delta$-functions (their arguments etc.).

### Singular manifolds  2.11

Strictly speaking, in the context of asymptotic expansions one deals with the singularities of the formal expansions of the product $G(p,\kappa)$ in $\kappa$. Note, however, that the structure of those singularities is the same (up to the strength of singularities) for all terms of the formal expansion. This is because the singularities are due to denominators, and the terms resulting from expansion of a factor have the same expression raised to different powers in the denominator. Therefore, it is convenient first to ignore the "dynamic" aspect of the problem (i.e. the fact that one deals with expansions) and consider the problem as "static" (i.e. existence of products only). After that we will see what modifications are needed to take into account approximation properties etc.

The singularities of the product $G(p)$ are due to the fact that each of the denominators may have zeros, and the singularities of individual factors in the space $P$ may overlap in a non-trivial fashion. One wishes to study such overlapping singularities based on the information about the individual factors.

---

the preferred choice). Then the task of recasting the final answers into an explicitly convergent form is postponed till the end of the construction — cf. [4], [5] where such an approach was motivated by calculational convenience of final formulas in a situation where the subtraction procedure is understood well enough from analogy with the UV renormalization in coordinate representation (cf. [36], [12]). Such an approach, however, may not be always desirable, especially in the non-Euclidean problems where the multi-level recursion structure of the problem with secondary expansions obscures the connection between various types of divergences that occur at intermediate steps. Moreover, the dimensional regularization may not always work in non-Euclidean situations [14].

---

The singularities of the $g$-th factor are localized on the manifold $\pi_g \subset P$ described by the equation $D_g(p) = 0$ :

$$\pi_g = \{p | D_g(p) = 0\} . \qquad 2.12$$

The singularity corresponding to the $g$-th factor is referred to as *elementary singularity*.

Note that $\pi_g$ may be a smooth manifold everywhere in $P$ (e.g. the cases 2.5 with $m_g \neq 0$ and 2.6). On the other hand, the points where $l_g(p) = 0$ for 2.5 with $m_g = 0$ (the so-called soft singularity) correspond to the apex of the light cone where the light cone is not a smooth manifold. More generally, such special singularities occur whenever the gradient nullifies, i.e. $\nabla D_g(p) = 0$ , where the gradient operator is

$$\nabla \equiv \frac{\partial}{\partial p} \equiv \left( \frac{\partial}{\partial p_1^\mu}, \frac{\partial}{\partial p_2^\mu}, \ldots \right). \qquad 2.13$$

If $D_g(p_0) = 0$ but $\nabla D_g(p_0) \neq 0$ , then there is a small neighbourhood $\mathcal{O}$ of $p_0$ in which the local coordinates $p = p(x, y, \ldots)$ may be chosen so that $D_g(p(x, y, \ldots)) = x$ . Then the factor becomes

$$\Delta_g(p(x, y, \ldots)) = (x - i0)^{n_g} . \qquad 2.14$$

We call the distributions that can be locally reduced to such form *Sohotsky-type distributions*. Their integrals with test functions are well-defined despite the fact that they are divergent by power counting (see also Sec. 3).

In this paper we consider only intersections of causal singularities. Formally, this means that all $\pi_g$ are smooth manifolds within $\mathcal{O}$. A sufficient condition to ensure that is to require that

$$\nabla D_g(p) \neq 0, \ \forall p \in \mathcal{O}, \ \forall g . \qquad 2.15$$

This means that we exclude from consideration the soft singularities — their study requires additional techniques that would lead us outside the scope of the present paper. The restriction will remain valid throughout this paper.

### Conventions for the gradient $\nabla = \partial/\partial p$  2.16

Geometric representation of the gradient vector $\nabla D(p)$ in the space $P$ of independent variable $p$ depends on how one chooses the scalar product in $P$. This is because the gradient is, strictly speaking, a vector from the co-space $P'$ whose embedding into $P$ is a matter of convention. However, the properties we are going to consider are expressed via relations that are linear in $\nabla$ (e.g. $\nabla D(p) \neq 0$ ) and are, therefore, unaffected by the conventions adopted. This allows some flexibility of notation depending on specific purposes: To build insight into the nature of intersections of Sohotski-type distributions, we will consider simple examples, in which it is convenient to regard $P$ is a Euclidean space. Then the gradient vector $\nabla D(p_0)$ is normal to the singular manifold $D(p) = 0$ . But after the general theorems are understood and we turn to concrete examples in the context of Feynman diagrams and Minkowski space, it is convenient to proceed in a formal algebraic manner. In particular, $p$ is then an array of Minkowskian vectors (the collection of loop momenta) and is of a mixed Euclidean/Minkowski type, e.g. $p \cdot p' = \sum_i p_i^\mu p'_{i\mu}$. Then the gradient $\nabla D(p_0)$ may happen to be tangential to the corresponding manifold $D(p) = 0$ (for instance, if $p$ is a Minkowskian 4-vector and $D(p) = p^2$ then $\nabla D(p_0) = 2p_0^\mu$). But that is a purely formal fact of no analytical consequence.

### Singular manifolds and singular subgraphs    2.17

Consider an arbitrary subset of lines, $\gamma \subset G$, and define:

$$\pi_\gamma \equiv \bigcap_{g \in \gamma \subset G} \pi_g. \qquad 2.18$$

In other words, $\pi_\gamma$ is the collection of points through which pass the singularities of each factor $g \in \gamma$. It is called *singular manifold* corresponding to the subproduct $\gamma$.

As in the Euclidean case, one may have $\pi_\gamma = \pi_{\gamma'}$ even when $\gamma \neq \gamma'$. However, the analytical nature of the singularity of the entire product $G(p)$ on $\pi_\gamma$ is determined by all the elementary singularities $\pi_g \supset \pi_\gamma$ and by all factors $g(p)$ that are singular on the entire $\pi_\gamma$, i.e. $\pi_g \supset \pi_\gamma$. So, whenever one deals with a singular manifold $\pi_\gamma$, the corresponding $\gamma \subset G$ contains *all* $g$ such that $\pi_g \supset \pi_\gamma$. The subproduct

$$\gamma(p) = \bigcap_{g \subset \gamma} g(p) \qquad 2.19$$

that contains all the factors from $G(p)$ that are singular on $\pi_\gamma$ is described as *complete*. The singular manifold $\pi_\gamma$ together with the associated complete subproduct γ will be called *singular subgraph* although at this stage we do not associate any graphical image with it.

The collection of all singular subgraphs of *G* is denoted as $S[G]$.

### Osculating singularities    2.20

In the Euclidean case, there were two basic geometric patterns of intersections of elementary singularities: factorizable and non-factorizable. Because now singular manifolds are non-linear, an additional pattern has to be dealt with, namely, osculating[i] singularities.

Let $g(p)$ be an intersection point of several non-linear manifolds $\pi_g$, $g \in \gamma$, and let $\mathcal{O}$ be its small neighbourhood. One can always find a smooth deformation of coordinates $p \to p'$ within $\mathcal{O}$ that transforms any one of the singular manifolds $\pi_g$ into a linear manifold so that the corresponding factor in the product takes the form $(v_g p' - i0)^{-n_g}$ where $v_g$ is a constant vector. There are two cases possible:

• A general case is when *all* the factors $g(p) \in \gamma$ that are singular at $p_0$ can be simultaneously linearized.

• When a simultaneous linearization is not possible, one deals with *osculating singularities*.

### Example: self-energy    2.21

Consider the integrand of a one-loop scalar self-energy insertion:

$$\frac{1}{-p^2 + m_1^2 - i0} \times \frac{1}{-(p-q)^2 + m_2^2 - i0}, \qquad 2.22$$

where $m_i \neq 0$, $p$ is a 4-dimensional integration momentum while $q$ is the external momentum. The integral of 2.22 over $p$ (after appropriate UV renormalization) is a function of $q^2$ and the masses. The interesting asymptotic regimes are as follows:

(i) $q^2 \to -\infty$ (the Euclidean regime),

(ii) $q^2 \to +\infty$ (the quasi-Euclidean regime; cf. Sec. 5.9);

(iii) $q^2 \to (m_1 + m_2)^2$ (the threshold regime).

To study (i), one chooses $q$ to be purely space-like and then Wick-rotates $p$ into the Euclidean region. Then the formalism of [4] becomes applicable.

The case (ii) is more interesting because from the point of view of momentum space, the picture here is much different from the Euclidean case. Indeed, choose $q$ to be purely time-like, $q = \left(\sqrt{q^2}, 0, 0, 0\right)$. An immediate Wick rotation is prohibited. Rescale the dimensional parameters, $p \to p(q^2)^{1/2}$ and $m_i^2 \to m_i^2 q^2$, so that the regime (ii) is replaced by $m_i \to 0$ for $q$ fixed, with both masses going to zero at the same asymptotic rate. Within the framework of the theory of *As*-operation, one constructs the corresponding expansion of 2.22 in the sense of distributions. The first step of the construction is to identify the singularities of the formal Taylor expansion, whose terms have the following general form:

$$\frac{1}{(-p^2 - i0)^{N_1}} \times \frac{1}{(-(p-q)^2 - i0)^{N_2}}. \qquad 2.23$$

The singularities are localized on two light cones, one shifted with respect to the other by $q$ as shown in 2.24:

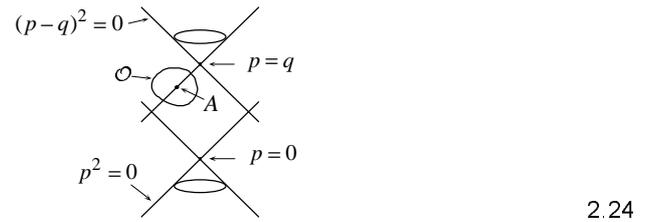

2.24

Two counterterms will be needed for the singularities at the apices of the light cones. Recall that in the Euclidean regime (i) one would have to deal with these two points only. In the present case, however, there are also singularities localized on the light cones away from their apices. Thus, in a small neighbourhood $\mathcal{O}$ of a general point $A$ of any of the light cones the local coordinates can be chosen so that the corresponding factor becomes $(x - i0)^{-N_i}$, which is non-integrable by power counting for $N_i$ large enough. The prescriptions of *As*-operation would require counterterms to be introduced for such singularities. However, it turns out that such light-cone singularities do not contribute to the final answer. The usual arguments explaining this rely on deformations of the contour of integration over $p^0$ into the complex plane. But for reasons discussed in the Introduction we would like to find an explanation for the absence of the counterterms in such cases directly in real Minkowskian momentum space.

---

[i] i.e. "kissing" (from Latin).

The threshold regime (iii) is the most difficult one. Indeed, the small expansion parameter $\kappa$ can be introduced by representing $q$ as follows:

$$q = (m_1 + m_2 + \kappa, 0,0,0) .$$  2.25

Then the expansion in $\kappa \to 0$ is equivalent to the expansion in $q^2 \to (m_1 + m_2)^2$.

A typical term of the formal expansion of 2.22 in $\kappa$ has the following structure of denominators that determine the structure of singularities:

$$\frac{1}{[-2m_1 r^0 - (r^0)^2 + \mathbf{r}^2 - i0]}$$
$$\times \frac{1}{[+2m_2 r^0 - (r^0)^2 + \mathbf{r}^2 - i0]^N} ,$$  2.26

where we have introduced the variable $r$:

$$p \equiv (p^0, \mathbf{p}) = (m_1 + r^0, \mathbf{r}) .$$  2.27

The singularities of the two factors are localized on two hyperboloids which touch (osculate) at the point $r = 0$ ($p = \bar{q} \equiv q|_{\kappa=0}$).

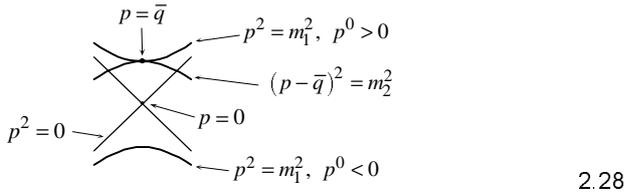

2.28

By analogy with the preceding example one has to conclude that the general points of the hyperboloids need not be supplied with counterterms when the expansion in the sense of distributions is constructed. But the argument does not work for the point $r = 0$ ($p = \bar{q}$). On the other hand, it is well-known from studies of analyticity properties of Feynman diagrams that the asymptotic regime under consideration corresponds to the point of non-analyticity; therefore, the expansion must contain a non-analytic term. In view of the fact that non-analytic dependences are contained in the coefficients of counterterms of the *As*-operation (cf. [4]), one concludes that it is exactly the counterterm for the singularity at $r = 0$ ($p = \bar{q}$) that should contain the non-analytic threshold behaviour.

Consider, however, the expansion in $\kappa$ defined as follows:

$$q = (m_1 - m_2 + \kappa, 0,0,0) .$$  2.29

Then there will also be a point where the two hyperboloids touch (cf. 2.30; only the osculating upper hyperboloids are shown; the dotted line shows the position of the unshifted hyperboloid ).

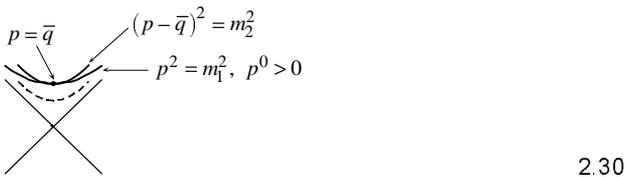

2.30

But according to the known analyticity properties, the expansion in $\kappa$ near 2.25 should not contain non-anaytic terms. This means that, unlike the preceding case, the osculation point of the two hyperboloids does not require a non-trivial counterterm.

### Example: collinear singularities 2.31

Another rather typical geometrical pattern of intersection of causal singularities occurs in the following product of two massless propagators:

$$\frac{1}{[-p^2 - i0]^{N_1}} \times \frac{1}{[-(p-q)^2 - i0]^{N_2}} ,$$  2.32

where $q^2 = 0$ (both $p$ and $q$ are 4-dimensional Minkowskian vectors). Such a product may have resulted from a formal expansion in one of the high-energy regimes (cf. [35]). The singularities of the two factors are localized on the two light cones as shown in 2.33.

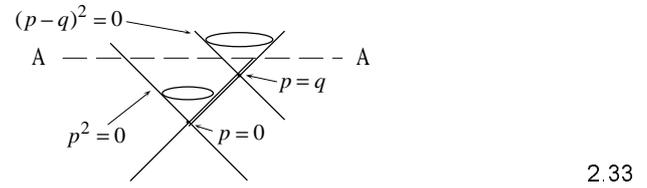

2.33

The first factor is singular on the light cone $p^2 = 0$ while the second, on a similar light cone whose apex is shifted to the point $p = q$. The two light cones intersect over the line

$$L = \{p | p = zq\}$$  2.34

(i.e. $p$ should be collinear to the light-like momentum $q$ which is the reason why this is called *collinear singularity*). Note that, in general, an intersection of two 3-dimensional manifolds is 2-dimensional, while in the present example it is 1-dimensional. To see what happens here, consider a cross section of the space of $q$ with a hyperplane $p^0 = \text{const} > q^0$ (represented by the line AA in 2.33) as in 2.35.

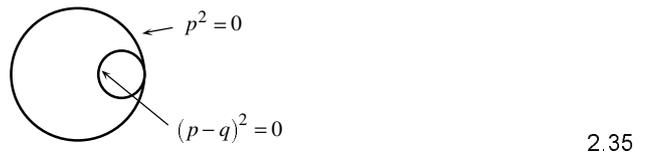

2.35

The intersections of the two light cones with this hyperplane are spheres that osculate at a point. When the cross section shifts along $p^0$, the spheres inflate/deflate and shift, but there is always one osculation point. If $0 < p^0 < q^0$ then the two spheres are side-by-side, otherwise one is inside the other; the two cases correspond to pinched and non-pinched singularity.

## One-dimensional case 3

Within the formalism of asymptotic operation, all problems are recursively reduced to investigation of one-dimensional singularities. Therefore, it is convenient to review the properties of the relevant one-dimensional distributions and remind ourselves of some basic properties of Fourier transforms that will be needed in the discussion of the multidimensional case.

### The Sohotsky distribution 3.1

Let $p$ be a one-dimensional integration variable running from $-\infty$ and $+\infty$. Recall the familiar one-dimensional Sohotsky distribution

$$\frac{1}{p-i0} \stackrel{\text{def}}{=} \lim_{\eta \to 0} \frac{1}{p-i\eta}, \qquad 3.2$$

which is well-defined despite the fact that the singularity is logarithmically divergent by power counting. As usual, the limit should be understood in the sense of the distribution theory, i.e. the above expression is not expected to be interpreted otherwise than in integrals with test functions $\varphi(p)$ (that are smooth everywhere and non-zero only in a compact region):

$$\int_{-\infty}^{+\infty} dp\, \varphi(p) \frac{1}{p-i0} \stackrel{\text{def}}{=} \lim_{\eta \to 0} \int_{-\infty}^{+\infty} dp\, \varphi(p) \frac{1}{p-i\eta}. \qquad 3.3$$

Note that both the integral and the limit on the r.h.s. exist for any such $\varphi(p)$.

The well-known Sohotsky formula reads:

$$\frac{1}{p-i0} = \text{PV} \frac{1}{p} + i\pi \delta(p), \qquad 3.4$$

where the first term is defined with the usual principal value prescription.

### Products of Sohotsky distributions 3.5

The next point is existence and non-existence of products of Sohotsky distributions. For instance, the product of the Sohotsky distribution and its complex conjugate,

$$\frac{1}{p-i0} \times \frac{1}{p+i0}, \qquad 3.6$$

does not exist as a distribution defined on all test functions unlike the square

$$\frac{1}{p-i0} \times \frac{1}{p-i0}, \qquad 3.7$$

which does.

In the context of analytical properties of Feynman diagrams (cf. e.g. [17]), this phenomenon is well-known and described in terms of pinched (Eq. 3.6) vs non-pinched (Eq. 3.7) integration contours. Indeed, if one integrates Eq. 3.7 with a function that is analytical in a neighbourhood of $p=0$, then the Cauchy theorem allows one to simply deform the contour of integration into the upper complex halfplane near $p=0$ because the poles of the two factors are on the same side of the contour; then the integrand will have no singularities at all on the integration contour. In the case of 3.6 this is not possible — the contour is "pinched" by the two poles.

### Interpretation in terms of Fourier transforms 3.8

Following, essentially, Hörmander (see [23] and refs. therein), consider the products 3.6 and 3.7 from the point of view of Fourier transformation. Then existence of products of functions that are singular at $p=0$ is determined by the behaviour of Fourier transforms at infinity. Such a behaviour may be different in different directions, and under certain conditions this may result in a well-defined product even if the factors are singular.

First recall that Fourier transforms of test functions are smooth functions that exhibit rapid decrease in any direction (i.e. decrease faster than any inverse power of their argument). However, Fourier transforms of distributions need not be decreasing. On the other hand, they may behave badly in some directions, and exhibit a rapid decrease in others. For instance, by straightforward integration one sees that

$$\frac{1}{p-i0} = i \int_{-\infty}^{+\infty} dx\, e^{-ikx} e^{-0x} \theta(x>0). \qquad 3.9$$

The Fourier transform $\theta(x>0)$ remains constant as $x \to +\infty$, but decreases rapidly as $x \to -\infty$.

For the complex-conjugate distribution the pattern of dangerous and non-dangerous directions is mirrored.

Note that raising the l.h.s. of 3.9 to an arbitrary power $n > 0$ results in a power of $x$ on the r.h.s.:

$$\left(\frac{1}{p-i0}\right)^n = i \int_{-\infty}^{+\infty} dx\, e^{-ipx} (-ix)^{n-1} e^{-0x} \theta(x>0). \qquad 3.10$$

The important point is that the growth of the integrand corresponding to a power singularity is no more than polynomial.

Further, consider the integral of 3.7 against a test function $\varphi(p)$. Using 3.9, we find:

$$\int_{-\infty}^{+\infty} dp \frac{1}{p-i0} \times \frac{1}{p-i0} \varphi(p)$$

$$= \text{const} \int_{-\infty}^{+\infty} dx \int_{-\infty}^{+\infty} dx'\, \theta(x>0)\, \theta(x'>0)\, \tilde{\varphi}(x+x'). \qquad 3.11$$

Thus, the convergence of the integral on the l.h.s. around $p=0$ has transformed into convergence of the integral on the r.h.s. at infinity in various directions. (Since the Fourier transform $\tilde{\varphi}(x)$ is a smooth function, integration over finite regions is not dangerous.) One sees that the $\theta$-functions ensure convergence in all directions other than those in the positive quadrant $x > 0, x' > 0$. However, convergence in those directions is ensured by the properties of $\tilde{\varphi}(x)$, as explained above.

One concludes that the integral 3.11 exists for any test function, therefore, the product 3.7 is a well-defined distribution.

On the other hand, a similar calculation for the product 3.6 gives:

$$\int_{-\infty}^{+\infty} dp \frac{1}{p-i0} \times \frac{1}{p+i0} \varphi(p)$$

$$\propto \int_{-\infty}^{+\infty} dx \int_{-\infty}^{+\infty} dx'\, \theta(x>0)\, \theta(x'<0)\, \tilde{\varphi}(x+x'). \qquad 3.12$$

One can see that there are directions in the plane $(x, x')$ — e.g. $x = \Lambda + x_0$, $x' = -\Lambda + x_0'$ — in which the $\theta$-functions are both equal to 1 while $\tilde{\varphi}(-x+x') = \tilde{\varphi}(-x_0+x_0') = \text{const}$ as $\Lambda \to +\infty$. The integral on the r.h.s., therefore, cannot be con-vergent for all test functions, and the product 3.6 is not a well-defined distribution.

As a further excercise, replace the second factor by, say, $\delta(p)$. Then the second $\theta$-function on the r.h.s. of 3.11 is replaced by a constant, and again, the integral does not exist for some test functions: the product of a $\delta$-function and a Sohotsky-type distribution is ill-defined.

On the other hand, if one of the distributions is a function that has any number of derivatives in a small neighbourhood around $p = 0$, then there are no dangerous directions associated with it at that point, so that its product with any distribution exists around $p = 0$.

### "Natural" existence 3.13

It is important to realize that *any* definition of an integral in infinite bounds contains — explicitly or implicitly — some sort of a cutoff, say, $|x| < \Lambda$, $\Lambda \to \infty$. Such a cutoff plays the role of a natural regularization. Then the absolute convergence of the integral 3.11 (established e.g. by power counting) implies that whatever (correct) regularization method one uses to give meaning to the product of the two Sohotsky distributions on the l.h.s. of 3.11, the limiting value for the integral on the l.h.s. will exist for any test function and it will be the same for all regularizations. This gives a precise meaning to the phrase that the product in such a case is "naturally defined" in the sense of distributions.

### Asymptotic expansions 3.14

### A single distribution 3.15

As a warm-up excercise, consider the distribution

$$\frac{1}{p + m - i0} \qquad 3.16$$

where $m$ is a real parameter. Its expansion in the sense of distributions as $m \to 0$ is:

$$\text{Eq. 3.16} \underset{m \to 0}{\approx} \sum_{n \geq 0} \frac{(-m)^n}{(p - i0)^{n+1}} \qquad 3.17$$

PROOF. It is sufficient to consider only the case $n = 0$. As usual, the expansion is sought in the form

$$\frac{1}{p + m - i0} = \frac{1}{p - i0} + c(m)\delta(p) + o(1), \qquad 3.18$$

and the coefficient $c(m)$ is determined from a consistency condition (in a manner similar to the Euclidean case; see sec. 7.4 of [4]):

$$c(m) = \int_{-\infty}^{+\infty} dp \left[ \frac{1}{p + m - i0} - \frac{1}{p - i0} \right] + o(1) = 0 + o(1). \qquad 3.19$$

The integration can be performed using the explicit representation 3.4 and symmetric cutoffs, $|p| < \Lambda$.

The expansion 3.17 can be regarded as the Taylor expansion of a translation along the $p$-axis. This explains the absence of non-trivial counterterms with logarithmic dependence on $m$: an infinitesimal translation involves differentiation, and differentiations in $p$ of the distribution $(p \pm i0)^{-n}$ are performed in a straightforward manner.

### Two factors 3.20

Consider a product of two factors like 3.16 with their singularities separated by a distance $o(m)$. Then one has the following expansion:

$$\frac{1}{p - i0} \times \frac{1}{p + m - i0} \underset{m \to 0}{\approx} \sum_{n \geq 0} \frac{(-m)^n}{(p - i0)^{n+2}}. \qquad 3.21$$

It can be proved in a way similar to the preceding example by using the procedure described in sec. 7 of [4]. Another way is to use partial fractioning,

$$\frac{1}{p - i0} \times \frac{1}{p + m - i0} = \frac{1}{m}\left(\frac{1}{p - i0} - \frac{1}{p + m - i0}\right), \qquad 3.22$$

and reduce this example to the case of one factor already considered.

Note that if one changes the sign of the imaginary parts of one of the factors, then non-trivial $m$-dependent counterterms appear in the expansion. For example:

$$\frac{1}{p + i0} \times \frac{1}{p + m - i0} \underset{m \to 0}{\approx} \sum_{n \geq 0} \frac{(-m)^n}{(p - i0)^{n+2}} - \frac{2\pi i}{m}\delta(p). \qquad 3.23$$

The simplest way to obtain this result is via a trick — switching the sign of imaginary part of the first factor using the Sohotsky formula 3.4. However, one can also derive it in a systematic way as in the preceding examples.

The conclusion is that the expansion 3.21 looks as if obtained by simply taking a product of the first factor (whose expansion in $m$ is trivial) and the expansion of the second factor, Eq. 3.17. Each term of such a product is a well-defined distribution. However, strictly speaking, existence of such a product does not guarantee that it will possess the necessary approximation properties to be the asymptotic expansion of the initial $m$-dependent product (the l.h.s. of 3.21). But it turns out that it does. Let us understand why this is so using the techniques of Sec. 3.8: in a general multidimensional situation, explicit calculations similar to those above may not be possible.

### Interpretation in terms of Fourier transforms 3.24

Similarly to 3.11, one has:

$$\int_{-\infty}^{+\infty} dp\, \varphi(p) \frac{1}{p - i0} \times \frac{1}{p + m - i0}$$
$$\propto \int_{-\infty}^{+\infty} dx \int_{-\infty}^{+\infty} dx'\, \theta(x > 0)[\theta(x' > 0)e^{-imx'}]\tilde{\varphi}(x + x'). \qquad 3.25$$

If one performs the Taylor expansion in $m$ on the r.h.s., the factor $\exp(-imx')$ (which is bounded by 1 everywhere) transforms into a polynomial of $x'$. The remainder term is then bounded by something like $O(m^N x'^N)$ which exhibits a polynomial growth as $x' \to \pm\infty$ — but not faster than polynomial. Therefore, the pattern of behaviour at infinity (polynomial growth *vs* rapid decrease) is the same both for non-expanded factors, the expanded factors, and the remainder term. This allows termwise integration of the product of expansions with preservation of approximation properties.

One sees that the "natural" existence (in the sense of Sec. 3.13) of the product of expansions of individual factors implies that such a product is indeed a correct asymptotic expansion for the initial non-expanded product; in other words, taking product and performing expansion are two commuting operations in such a case.

The above examples provide enough motivation for analogous multidimensional constructions to which we now turn.

## Products of Sohotsky distributions in $D > 1$     4

We are going to generalize the notion of dangerous directions to the multidimensional case. We first consider the case of linear denominators (Sec. 4.1) and derive a simple geometric criterion of existence of products of Sohotsky-type distributions. We then consider smooth deformations of coordinates (Sec. 4.14) and see what form the criterion takes. The notion of singular wave front emerges naturally (Sec. 4.23) and allows one to easily obtain a criterion (essentially due to Hörmander; cf. Theorem IX.45 in [23]) for the general non-linear case including the so-called osculating singularities (Sec. 4.34). Finally, the extension of the criterion to the case of asymptotic expansions of Sohotsky-type distributions becomes straightforward (Sec. 4.41).

### Linear case     4.1

### Two factors     4.2

Consider a product of two Sohotsky-type distributions integrated against a test function $\varphi(p)$ localized in the region $\mathcal{O}$:

$$\int \mathrm{d}p \, \varphi(p) \frac{1}{(v_1 p - i0)(v_2 p - i0)} \,. \quad 4.3$$

For simplicity of notation, we do not consider general powers explicitly. The distributions with $+i0$ can be taken into account by changing signs of (some of) $v_i$.

One can see that the product is a well-defined distribution unless the two vectors $v_1$ and $v_2$ look in opposite directions. Indeed, if they are linearly dependent then the problem effectively degenerates into a 1-dimensional one: if they look in the same direction then the product is a well-defined distribution; if they look in opposite directions then the product is non-integrable. If the two vectors are linearly independent then one deals with a direct product of two distributions that are well-defined separately, and such a product is a well-defined distribution too.

### Three factors     4.4

Under the same assumptions, consider the product of three Sohotsky-type distributions:

$$\int \mathrm{d}p \, \varphi(p) \frac{1}{(v_1 p - i0)(v_2 p - i0)(v_3 p - i0)} \,. \quad 4.5$$

The singularities of the $g$-th factor are localized on the linear manifold

$$\pi_g = \{ p \mid v_g p = 0 \} \,. \quad 4.6$$

First, without loss of generality one assumes that the origin is the only point of intersection of the three manifolds. Second, one has to assume that the product of each pair of the three factors is well-defined as a distribution everywhere. A standard reasoning [4] shows that the product 4.5 is then integrable with any test function that is equal to zero in any small neighbourhood of the origin of coordinates. It remains to study the behaviour of the product near that isolated point.

### Existence of the product: analytical criterion     4.7

Using the representation

$$\frac{1}{A - i0} = i \int_0^{+\infty} \mathrm{d}\lambda \, e^{-i\lambda A - 0\lambda} \,, \quad 4.8$$

one can rewrite 4.5 as follows:

$$\text{Eq. 4.5} = i^3 \int_0^{+\infty} \mathrm{d}\lambda_1 \int_0^{+\infty} \mathrm{d}\lambda_2 \int_0^{+\infty} \mathrm{d}\lambda_3 \, \tilde{\varphi}(v_1\lambda_1 + v_2\lambda_2 + v_3\lambda_3) \,, \quad 4.9$$

where $\tilde{\varphi}(x)$ is the Fourier transform of $\varphi(p)$:

$$\tilde{\varphi}(x) \propto \int \mathrm{d}p \, e^{-ipx} \varphi(p) \,. \quad 4.10$$

One can see that the rapid decrease of $\tilde{\varphi}$ makes the integral convergent at large $\lambda_g$ in any directions if the following condition is satisfied

$$v_1\lambda_1 + v_2\lambda_2 + v_3\lambda_3 \neq 0, \quad \text{for all } \lambda_g > 0 \,. \quad 4.11$$

Then the product in the integrand of 4.5 exists in the sense of distributions.

Note that for, say, $\lambda_3 = 0$, the first inequality in 4.11 degenerates into the condition of existence of the product of the first two factors. Therefore, instead of making the assumption of existence of subproducts one can simply require that the inequality in 4.11 hold for all $\lambda_g \geq 0$, $\sum_g \lambda_g > 0$.

### Geometric criterion     4.12

Consider the set of all possible linear combinations $v_1\lambda_1 + v_2\lambda_2 + v_3\lambda_3$ with $\lambda_g > 0$. We call it _proper cone_ if it lies strictly within one half of the linear space of $v_i$:

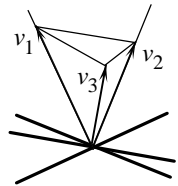

    4.13

In such a case the integral 4.9 and the product 4.5 are well-defined.

The generalization to more than three factors is straightforward.

Note that the property of a set of vectors to span a proper cone is invariant under arbitrary linear transformations.

### Non-linear deformations of coordinates     4.14

As a next step towards the general case, notice that existence of a distribution is a fact that survives any *non-linear* smooth mapping of coordinates. To rewrite the above criterion in a form that is "covariant" under such transformations, consider a neighbourhood $\mathcal{O}$ of the origin and perform a smooth ($C^\infty$) deformation of the coordinates so that

$$v_g p \to v_g p + O(p^2) \equiv D_g(p) \,. \quad 4.15$$

Fig. 4.13 is modified as follows:

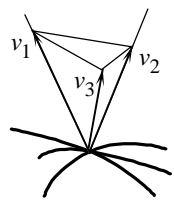

    4.16

Then Eq. 4.5 becomes:

$$\int_{\text{supp}\,\varphi \subset \mathcal{O}} dp\, \varphi(p) \frac{1}{(D_1(p) - i0)(D_2(p) - i0)(D_3(p) - i0)} \quad . \qquad 4.17$$

The singularities of the three factors are now localized on non-linear manifolds described by non-linear equations:

$$\pi_g = \{p | D_g(p) = 0\}\,. \qquad 4.18$$

The integration measure is transformed with a Jacobian that is smooth within $\mathcal{O}$ and it is absorbed into the redefined test function. The region $\mathcal{O}$ is deformed accordingly.

Moreover, the choice of the origin of coordinates cannot affect existence of the product, so the origin can be shifted arbitrarily. Denote the point of intersection of the three manifolds 4.18 as $p_0$. Since $v_g = \nabla D_g(p_0)$, the criterion 4.11 is rewritten as follows:

$$\lambda_1 \nabla D_1(p_0) + \lambda_2 \nabla D_2(p_0) + \lambda_3 \nabla D_3(p_0) \neq 0 \text{ for all } \lambda_g > 0\,. \quad 4.19$$

### Criterion for any number of factors 4.20

Consider now a product containing an arbitrary number of Sohotsky-type factors:

$$\prod_{g \in \gamma} \frac{1}{(D_g(p) - i0)^{n_g}} \quad . \qquad 4.21$$

Without loss of generality one can assume that there is only one point $p_0$ (not a higher-dimensional manifold) where the singular manifolds $\pi_g$ intersect all at once; otherwise by a smooth change of coordinates one can deform the intersection $\pi_\gamma \equiv \bigcap_{g \in \gamma} \pi_g$ into a linear subspace (within $\mathcal{O}$), and then only the coordinates that are transverse to $\pi_\gamma$ will matter.

By analogy with the preceding examples, one writes down the following criterion of existence of the product 4.21 at $p_0$:

$$\sum_{g \in \gamma} \lambda_g \nabla D_g(p_0) \neq 0 \quad \text{for all} \quad \lambda_g \geq 0\,, \quad \sum_{g \in \gamma} \lambda_g > 0\,. \quad 4.22$$

If all subproducts are exist, then it is sufficient to consider only $\lambda_g > 0$.

The heuristic meaning of 4.22 is the same as in the example above: as is discussed in 4.27 the vector $\nabla D(p_0)$ describes the dangerous direction for the distribution $(D(p) - i0)^{-n}$ at its singular point $p_0$, i.e. the only direction in which its Fourier transform does not decrease rapidly, thus jeopardizing convergence in that direction of Fourier-transformed integrals with test functions.

Recall that the above reasoning allows one to prove the criterion only in situations where all the non-linear singularities can be "flattened" simultaneously by a smooth transformation to the form of 4.13. Nevertheless, it is useful to summarize the experience gained in a intuitive geometrical form.

### Wave front of a Sohotsky-type distribution 4.23

Consider a distribution $(D(p) - i0)^{-n}$ defined within a region $\mathcal{O}$. Its singularities are localized at the points where $D(p) = 0$. Such points constitute a manifold, $\pi = \{p | D(p) = 0\}$. Assume that $\pi$ is well-behaved (i.e. is non-

singular in the sense of differential geometry) within $\mathcal{O}$, which means that

$$\nabla D(p)|_{D(p) = 0} \neq 0\,, \qquad p \in \mathcal{O}\,. \qquad 4.24$$

Construct the <u>wave front</u> of the distribution $(D(p) - i0)^{-n}$ by attaching the vector $\nabla D(p_0)$ (describing the "dangerous" direction; cf. also Sec. 4.27 below) to each singular point $p_0 \in \pi$:

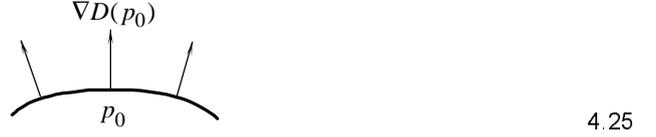

4.25

Now the criterion 4.22 can be reformulated in geometric terms:

### Geometric form of the criterion 4.26

For each point $p_0$ where some of the factors of the product 4.21 are singular, select all those and only those factors that are singular at that point. If the corresponding dangerous directions $\nabla D_g(p_0)$ span a proper cone — and if this holds true for any such point $p_0$ — then the product is integrable with arbitrary test functions localized within $\mathcal{O}$.

### Why $\nabla D(p_0)$ is a dangerous direction 4.27

The above analysis was based on a deformation of coordinates that made all singular manifolds flat simultaneously. This is not always possible (recall osculating singularities, Sec. 2.20). Therefore, it is useful to consider the microlocal mechanism that ensures existence of the products in terms of Fourier transforms without invoking global deformations. It is at this point that the notion of wave front becomes genuinely non-trivial.

Consider again the distribution $\Delta(p) = (D(p) - i0)^{-a}$ near a point $p_0$ where $D(p_0) = 0$ but $\nabla D(p_0) \neq 0$. It turns out that any direction $n$ other than the one described by $\nabla D(p_0)$ is not dangerous at $p_0$ in the following sense: one can find a test function $\psi(p)$ that is localized sufficiently closely around $p_0$ and such that the Fourier transform $\tilde\Delta_\psi(x) = \int dp\, e^{ixp} \psi(p) \Delta(p)$ decreases rapidly (i.e. faster than any power) in the direction of $n$, i.e. $\tilde\Delta_\psi(\lambda n) \to 0$ faster than any power $\lambda^{-N}$ as $\lambda \to +\infty$.

The proposition is quite intuitive: behaviour at infinity of Fourier transforms is determined by the behaviour of the original in an arbitrarily small neighbourhood of the origin. The denominators of the distributions we deal with are linear functions up to $O(p^2)$ terms which are — as the preceding examples suggest — negligible if one focuses on a sufficiently small neighbourhood, so that the distributions resemble the standard Sohotsky distribution that depends on the coordinate directed along $\nabla D(p_0)$. One may also note that multiplication by a smooth function does not change this property.

The formal proof which is a straightforward technical implementation of the above heuristic idea is given in Sec. 4.28 below.

A somewhat stronger version of the above property will be useful: the rapid decrease $\tilde\Delta_\psi(\lambda n) \to 0$ in $\lambda$ is uniform with re-spect to all directions $n$ outside some (arbitrarily narrow) cone $C'$ containing $\nabla D(p_0)$. This is established as follows. First, the proof of the above proposition allows one to obtain bounds that are, in fact, uniform with respect to all directions from a sufficiently narrow cone $C_n$ around $n$. Then one notices that directions can be represented as points of the unit sphere which is compact; it follows that for any cone $C'$ one can select a finite number of directions $n_i$ whose cones $C_{n_i}$ contain all directions outside $C'$. Choosing the largest bound among those corresponding to $C_{n_i}$, one obtains the required uniform one.

### Proof of the proposition of Sec. 4.27     4.28

Let the test functions $\psi(p)$ be localized within a small neighbourhood $\mathcal{O}$ of the point $p_0$. By a shift of coordinates one ensures that $p_0 = 0$. Consider the integral

$$\int dp\, e^{-i\lambda np} \psi(p) \frac{1}{(L(p)+i0)^a} \qquad 4.29$$

Split $p = (p_\perp, p_\parallel)$ where the first component is orthogonal to the manifold $L(p) = 0$ while the second, tangential to it at 0. The Taylor theorem guarantees that $L(p) = p \cdot \nabla L(0) + O(p^2)$ for $p \to 0$. Therefore, choosing $\mathcal{O}$ small enough allows one to find a smooth interpolating function $\Psi(p)$ such that $\Psi(p) = p \cdot \nabla L(0)$ for all $p$ sufficiently large, and such that $\Psi(p) = L(p)$ within $\mathcal{O}$. Without loss of generality, the following reasoning assumes $L(p) \equiv \Psi(p)$.

There are two cases to be considered:
A. $n$ is directed at an angle to $\nabla L(0)$;
B. The directions of $n$ and $\nabla L(0)$ are exactly opposite.

CASE A. Let us exhibit the fact that the oscillations of the exponential *along* the singular manifold $L(p) = 0$ for $\lambda \to +\infty$ suppress the integral as in the case of a smooth function — irrespective of the singular behaviour in the transverse directions. To this end, split $p_\perp = (p', p'')$ where $p$ is directed along the projection of $n$ onto the subspace of $p_\perp$, so that $p''$ is orthogonal to both $n$ and $\nabla L(0)$. In the coordinates $(p', p'', p_\parallel)$, one has $n = (n', 0, n_\parallel)$. Finally, choose new coordinates $(\kappa', \kappa'', \kappa_\parallel)$ so that:

$$\kappa_\parallel = \Psi(p), \quad n'\kappa' = n'p' - n_\parallel \kappa_\parallel, \quad \kappa'' = p'' . \qquad 4.30$$

One can see that the closer $n$ to $\nabla L(0)$, the smaller $n'$, and the closer $\psi(p)$ should be to $p \cdot \nabla L(0)$, which requires a smaller $\mathcal{O}$.

In the new coordinates, the integral becomes

$$\int d\kappa\, e^{-i\lambda(n_\parallel \kappa_\parallel + n'\kappa')} \psi(\kappa) \frac{1}{(\kappa_\parallel + i0)^a} . \qquad 4.31$$

The Jacobian of the coordinate transformation was absorbed into the test function $\psi(\kappa)$. The integrand is smooth in $\kappa'$ so that one can perform the standard integration-by-parts trick (Sec. 3.8) to establish the rapid decrease of the entire integral.

CASE B. Here one makes use of the fact that one deals with a "slightly deformed" Sohotsky distribution whose Fourier transform is a "slightly deformed" $\theta$-function with essentially similar nullification properties.

Similarly to case A, change the coordinates in 4.29 as follows: $\kappa_\parallel = L(p)$, $\kappa_\perp = p_\perp$. Denote $N = -n_\parallel > 0$. Then using 4.8 (with the integration variable changed to $\Lambda$), rewrite the integral as:

$$\int_{\lambda N}^{+\infty} d\Lambda \int d\kappa_\parallel d\kappa_\perp \psi(\kappa) \exp i\left[\Lambda \kappa_\parallel + \lambda N \Omega(\kappa) - \lambda n_\perp \kappa_\perp\right], \qquad 4.32$$

where $\Omega(\kappa) = p_\parallel(\kappa) - \kappa_\parallel$. If the neighbourhood $\mathcal{O}$ was chosen small enough, one can ensure that $|\Omega'(\kappa)| \le \delta < 1$. This allows one to integrate by parts:

$$\int d\kappa_\parallel e^{i\left[\Lambda \kappa_\parallel + \lambda N \Omega(\kappa)\right]} \{\ldots\}$$
$$= \int d\left(e^{i\left[\Lambda \kappa_\parallel + \lambda N \Omega(\kappa)\right]}\right) \frac{1}{\Lambda + \lambda N \Omega'(\kappa)} \{\ldots\} = \text{etc.} \qquad 4.33$$

The new factor does not cause problems because it is smooth and the denominator does not turn to zero. In this way one obtains bounds for the integral over $\kappa$ which, after integration over $\Lambda$, make manifest the rapid decrease of the integral. This completes the proof.

### Osculating singularities     4.34

This is the situation when the singular manifolds $D_g(p) = 0$ touch rather than intersect in a general fashion. We are going to show that the criterion obtained above works here too.

It is sufficient to consider the simplest case of 2-dimensional Euclidean $p = (p_1, p_2)$ and the product of two factors:

$$\Delta_1(p)\Delta_2(p), \quad \Delta_g(p) = (D_g(p) - i0)^{-1},$$
$$D_1(p) = p_1 - i0, \quad D_2(p) = p_1 - p_2^2 - i0. \qquad 4.35$$

The dangerous directions are shown in 4.36:

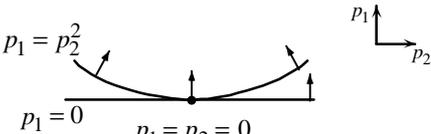

4.36

It is natural to suppose that since the quadratic terms played no role in Secs. 4.14–4.20 they should not be important here, too.

Because the singular manifolds do not intersect away from $p = 0$, it is sufficient to consider an arbitrarily small neighbourhood of that point.

The dangerous direction at $p = 0$, the same for both factors, is

$$\nabla D_1(0) = \nabla D_2(0) = (1,0) \equiv V . \qquad 4.37$$

Take a cone $C_{\text{sing}}$ of directions around $V$. Choose it narrow enough so that $C_{\text{sing}} + C_{\text{sing}}$ — which is the cone spanned by all possible pairs of vectors from $C_{\text{sing}}$ — is a proper cone.

Furthermore, according to the last paragraph of Sec. 4.27, given $C_{\text{sing}}$, one can find a pair of test functions $\psi_1(p)$ and $\psi_2(p)$ localized around $k = 0$ such that Fourier transforms $\tilde{\Delta}_g(x)$ of the distributions $\psi_g(p) \times \Delta_g(p)$ are rapidly and uniformly decreasing in all directions except those in the cone $C_{\text{sing}}$.

Because the functions $\psi_g(p)$ are smooth and non-zero in a small neighbourhood of $p = 0$, it is sufficient to prove existence of the product

$$\frac{\psi_1(p)}{D_1(p) - i0} \times \frac{\psi_2(p)}{D_2(p) - i0} \quad . \qquad 4.38$$

Consider the integral

$$\int dp_1 \, dp_2 \, \frac{\psi_1(p)}{D_1(p) - i0} \times \frac{\psi_2(p)}{D_2(p) - i0} \times \varphi(p) \qquad 4.39$$

for an arbitrary test function $\varphi(p)$ (whose localization properties are not important because the distributions themselves are localized as needed). Its representation in terms of Fourier transforms is as follows:

$$\int d^2x \, d^2y \, \tilde{\Delta}_1(x) \tilde{\Delta}_2(y) \tilde{\varphi}(x+y) \quad . \qquad 4.40$$

One sees that the only directions in the 4-dimensional space of the aggregate integration variable $(x, y)$ where the product $\tilde{\Delta}_1(x)\tilde{\Delta}_2(y)$ does not exhibit a rapid decrease are the ones that lie within the direct product $C_{\text{sing}} \times C_{\text{sing}}$, i.e. the directions of the form $(x, y) = \lambda(V_1, V_2)$ with $V_i \in C_{\text{sing}}$. However, the construction of $C_{\text{sing}}$ ensures that the integrand in such directions is suppressed by the rapid decrease of the test function. This proves absolute convergence of the integral 4.40 and, therefore, existence of the products 4.38 and 4.35.

The proof immediately generalizes to the case of any number of factors.

To summarize, the criterion of Sec. 4.20 is valid irrespective of the pattern of intersection of the singular manifolds $\pi_g$ — whether they intersect in a transverse fashion or osculate — as long as the other conditions are satisfied (most notably, smoothness of all manifolds $\pi_g$, or $\nabla D_g(p) \neq 0$ within $\mathcal{O}$).

### Extension to asymptotic expansions 4.41

One can immediately extend the above results to obtain a criterion of existence of asymptotic expansions (cf. Sec. 3.14). It will be sufficient to consider the case of two factors.

Let $\kappa$ be a parameter that goes to zero, $\kappa \to +0$. Consider a product of Sohotsky-type distributions $(D_g(p,\kappa) - i0)^{-n_g}$. Assume that for all $(p,\kappa) \in \mathcal{O} \times (-\varepsilon, \kappa_0)$ (with some $\varepsilon > 0$), $D_g(p,\kappa)$ are smooth in $(p,\kappa)$ and $\nabla D_g(p,\kappa) \neq 0$. Then if the product of $(D_g(p,0) - i0)^{-N}$ exists for arbitrarily large $N$ in the "natural" sense (i.e. in virtue of the Hörmander criterion)[i], then the asymptotic expansion in $\kappa \to 0$ in the sense of distributions of the original product is equal to the product of the asymptotic expansions of factors (recall that the latter expansions are simply the Taylor expansions).

The proof is a straightforward modification of the reasoning in Sec. 4.28 in the spirit of Sec. 3.14.

---

[i] which means that the product of Taylor expansions of factors is well-defined to all orders in $\kappa$.

## Applications 5

### Landau equations 5.1

The original Landau equations [16], [17] provided an answer to this question: at what values of external parameters — masses and momenta — does a given multiloop diagram cease to be an analytic function? The problem we are considering now is phrased differently: given an integrand of a multiloop diagram for some fixed values of external momenta and masses, where in the space of integration momenta does the integrand (the product of propagators) cease to be a well-defined distribution? The close relation of the two questions is clearly seen from the point of view of the theory of asymptotic operation which establishes a connection between the following two effects:

(i) non-analyticity of the integrated diagram as a function of external parameters at some values of the latter (i.e. a presence of, say, logarithmic terms in the corresponding asymptotic expansion near those values) and

(ii) non-integrability of the Taylor-expanded integrand (for a discussion see [1], [4]).

### Derivation 5.2

Consider a product of propagators of the form 2.2. It need not be a complete integrand — just a subproduct whose factors's causal singularities pass over a region $\mathcal{O} \subset P$. Consider the points where the causal singularities of all the factors from $G$ overlap. Such points are solutions of the system of equations

$$\{L_g(p) = 0\}_{g \in G} \quad . \qquad 5.3$$

However, at some of those points the product exists, and at others, it does not — as described by the criterion of Sec. 4.20. Recalling that $p$ is the collection of all loop momenta, $p = (p_1, p_2, \ldots p_l)$, and that the gradient operator is then represented as 2.13, one rewrites 4.22 as follows:

$$\sum_{g \in \gamma} \lambda_g \begin{pmatrix} \dfrac{\partial l_g^2(p)}{\partial p_1^\mu} \\ \vdots \\ \dfrac{\partial l_g^2(p)}{\partial p_l^\mu} \end{pmatrix} \neq 0 \quad \text{for all } \lambda_g \geq 0, \quad \sum_{g \in \gamma} \lambda_g > 0 \quad . \qquad 5.4$$

The points $p = (p_1, p_2, \ldots p_l)$ that satisfy 5.3 and 5.4 are the points where the product exists in a natural sense (as explained in Sec. 3.13) — such singularities are traditionally called *non-pinched*, while $p$ satisfying 5.3 but not 5.4 constitute a manifold in the integration space where the product does not exist — the manifold of *pinched* singularities.

The system of Landau equations is obtained if one replaces inequalities by equalities in 5.4, and reinterprets it as a description of pinched singularities rather than conditions for existence of the product.

### Non-covariant gauges 5.5

In the theory of non-covariant light-cone gauges ($nA = 0$ with $n^2 = 0$) one encounters propagators of the following form:

$$\frac{1}{p^2 + i0} \times \left( g_{\mu\nu} - \frac{n_\mu p_\nu + n_\nu p_\mu}{n\,p} \right), \qquad 5.6$$

where $n^2 = 0$, $n^0 > 0$. An important technical problem (cf. e.g. [31]) is the singular structure of Feynman diagrams involving such propagators in view of the additional singularity in 5.6. The so-called Mandelstam-Leibbrandt prescription (for a review see [34]) consists in using

$$n\,p \to n\,p + i0 \times \operatorname{sign} p^0. \qquad 5.7$$

Using the formalism developed above, it is not difficult to see why this prescription works:

Indeed, the singular manifolds corresponding to the two factors in the denominator of 5.6 are described by $p^2 = 0$ and $n\,p = 0$, and intersect over the line $p \propto n$. Consider any point $p$ on that line and evaluate the gradients:

$$\nabla_\mu p^2 = 2 p_\mu, \quad \nabla_\mu \left( \operatorname{sign} p^0 \times n\,p \right) = \operatorname{sign} p^0 \times n_\mu. \qquad 5.8$$

(The overall sign is switched so as to make the imaginary infinitesimal parts the same sign, as required by the formalism of the preceding sections.)

One sees that the two gradients are the same up to a positive coefficient — irrespective of whether $p$ is on the upper or lower half-cone. The criterion 4.26 then tells us that the product 5.6 with the prescription 5.7 is a well-defined distribution. This also means that the wave front of the propagator 5.6 on the light cone is not different from that of the simple scalar propagator. From the point of view of studying intersections of causal singularities of several such propagators, this means that there will be no new pinches etc. as compared with the covariant case. Of course, the propagator 5.6 has an additional singularity $n\,p = 0$ outside the light cone. But since such additional singularities are localized on a linear manifold, studying their intersections presents no difficulty.

A reasoning of this sort allows one to largely avoid explicit calculations when studying the singular structure of diagrams (cf. examples of such calculations in [31]).

### Quasi-Euclidean asymptotic regimes 5.9

As a last application, we consider the class of asymptotic regimes which, although not allowing Wick rotation, nevertheless result in the same formulas as in the Euclidean case. A simple example of such a regime was given in Sec. 2.21.

It is not difficult to understand (cf. Fig. 2.24) that no modifications (no new counterterms) to the formulas of Euclidean asymptotic operation are necessary as long as the following holds true in the asymptotic limit corresponding to a particular regime: Apices of light cones corresponding to various propagators should either merge in the asymptotic limit, or stay away from each other at a non-light-like separation; they are also not allowed to approach mass shell surfaces in the asymptotic limit. More precisely, if, as in [4], $m$, $q$ and $M$, $Q$ are small and large masses and external momenta of the problem, respectively (for definiteness, all the momenta enter the diagrams), and if one formulates the asymptotic regime in terms of $m, q \to 0$ with $M, Q$ fixed, then $q$ should vanish componentwise whereas $Q$ are allowed to be arbitrary non-lightlike provided the asymptotic configuration does not correspond to any of the thresholds. More precisely, in the context of a concrete 1PI diagram $G$, let $C_\alpha$ be a cut of the diagram with $Q_\alpha$ being the momentum entering any one of the two resulting components of $G$, and let $M_\alpha$ be the sum of masses of the lines cut by $C_\alpha$; further suppose that $Q_\alpha^2 \neq M_\alpha^2$ for any cut $\alpha$ in the asymptotic limit ($m = q = 0$). Then the asymptotic expansion is given by the same formulas as in the case of purely Euclidean asymptotic regimes [4], [5].

Finally, it should be emphasized that, strictly speaking, the pinch/non-pinched classification of the singularities can be interpreted in the context of asymptotic operation as equality to zero of counterterms at the corresponding points of singular manifolds. Therefore, one might proceed with the construction of non-Euclidean asymptotic operation ignoring such issues, and then rediscover nullification of counterterms for non-pinched singularities by an inspection of the resulting integrals. As a matter of fact, analytical study of the most complex types of non-Euclidean singularities (boundary points of pinched manifolds) may be psychologically easier if one ignores the pinch problem altogether. But the existing treatments *start* with the pinch/non-pinched classification (cf. [35]), so it would have been inconvenient to postpone establishing a connection with what is already known until the new theory is completely developed.

### Acknowledgments

I thank A. V. Radyushkin and G. Sterman for discussions of the problem of non-Euclidean asymptotic expansions.

Parts of this work were done during two visits to the Physics Department of Penn State University.

At different stages, this work was supported in parts by the CTEQ collaboration, the Russian Fund for Fundamental Research 95-02-05794, the International Science Foundation under grants MP9000/9300, and the U.S. Department of Energy under grant DE-FG02-90ER-40577.

I am indebted to P. Cherzor for pointing out several misprints.